\documentclass[
reprint,
superscriptaddress,
amsmath,amssymb,
aps,
pra,
floatfix,
]{revtex4-2}

\def\final{0}

\usepackage{braket}
\usepackage{mathtools}
\usepackage{graphicx}
\usepackage{color, xcolor}
\usepackage{dcolumn}
\newcolumntype{z}[1]{D{.}{.}{#1}}
\usepackage{physics}
\usepackage{upgreek}
\usepackage{subfigure}
\usepackage{tabularx}
\usepackage{hyperref}
\newcolumntype{Y}{>{\centering\arraybackslash}X}

\usepackage{amsthm}

\newtheoremstyle{italicheader}  
  {}                            
  {}                            
  {\rm}                         
  {}                            
  {}                            
  {.}                           
  { }                           
  {\itshape\thmname{#1}\thmnumber{ #2}\thmnote{ (#3)}} 

\theoremstyle{italicheader}
\newtheorem{theorem}{Theorem}

\usepackage[linesnumbered,ruled ]{algorithm2e}
\SetKwInput{KwOutput}{Output}
\SetKwInOut{Input}{Input}
\SetKwInput{Output}{Output} 
\UseRawInputEncoding 

\usepackage{xargs}
\usepackage[colorinlistoftodos,prependcaption]{todonotes}

\definecolor{Asparagus}{rgb}{0.53, 0.66, 0.42}
\definecolor{cornflowerblue}{rgb}{0.39, 0.58, 0.93}
\definecolor{darkolivegreen}{rgb}{0.33, 0.42, 0.18}
\definecolor{awesome}{rgb}{1.0, 0.13, 0.32}

\newtheorem{definitionenv}{Definition}
\newtheorem{lemmaenv}[definitionenv]{Lemma}

\newtheorem{corollaryenv}[definitionenv]{Corollary}
\newtheorem{propositionenv}[definitionenv]{Proposition}
\newtheorem{conjectureenv}[definitionenv]{Conjecture}
\newtheorem{remarkenv}[definitionenv]{Remark}
\newenvironment{remark}{\begin{remarkenv}\rm}{\end{remarkenv}}
\newcommand{\br}{\begin{remark}}
	\newcommand{\er}{\end{remark}}

\newtheorem{exampleenv}{Example}
\newtheorem{app-lemmaenv}[section]{Lemma}

\newenvironment{definition}{\begin{definitionenv}\rm}{\end{definitionenv}}
\newenvironment{lemma}{\begin{lemmaenv}\rm}{\end{lemmaenv}}
\newenvironment{corollary}{\begin{corollaryenv}\rm}{\end{corollaryenv}}
\newenvironment{example}{\begin{exampleenv}\rm}{\end{exampleenv}}
\newenvironment{proposition}{\begin{propositionenv}\rm}{\end{propositionenv}}
\newenvironment{conjecture}{\begin{conjectureenv}\rm}{\end{conjectureenv}}
\newenvironment{app-lemma}{\begin{app-lemmaenv}\rm}{\end{app-lemmaenv}}

\newcommand{\bd}{\begin{definition}}
	\newcommand{\ed}{\end{definition}}
\newcommand{\bl}{\begin{lemma}}
	\newcommand{\el}{\end{lemma}}
\newcommand{\elp}{\hspace*{\fill} $\Box$
\end{lemma}}
\newcommand{\bt}{\begin{theorem}}
\newcommand{\et}{\end{theorem}}
\newcommand{\etp}{\hspace*{\fill} $\Box$
\end{theorem}}
\newcommand{\bc}{\begin{corollary}}
\newcommand{\ec}{\end{corollary}}
\newcommand{\ecp}{\hspace*{\fill} $\Box$
\end{corollary}}
\newcommand{\bcj}{\begin{conjecture}}
\newcommand{\ecj}{\end{conjecture}}

\newcommand{\be}{\begin{example}}
\newcommand{\ee}{\end{example}}
\newcommand{\eep}{\hspace*{\fill} $\Box$
\end{example}}
\newcommand{\bp}{\begin{proposition}}
\newcommand{\ep}{\end{proposition}}
\newcommand{\epp}{
\end{proposition}}

\newcommand{\wt}{\mathrm{wt}}

\ifnum\final=0
\newcommand{\mynote}[2]{{\color{#1} \marginpar{\tiny #2}}}
\newcommand{\mybignote}[2]{{\color{#1} $\langle \langle$ #2$\rangle \rangle$}}
\newcommandx{\rednote}[2][1=]{\todo[linecolor=red,backgroundcolor=red!25,bordercolor=red,#1]{#2}}
\newcommandx{\bluenote}[2][1=]{\todo[linecolor=blue,backgroundcolor=blue!25,bordercolor=blue,#1]{#2}}
\newcommandx{\yellownote}[2][1=]{\todo[linecolor=yellow,backgroundcolor=yellow!25,bordercolor=yellow,#1]{#2}}
\newcommandx{\greennote}[2][1=]{\todo[inline,linecolor=olive,backgroundcolor=green!25,bordercolor=olive,#1]{#2}}

\else
\newcommand{\mynote}[2]{}
\newcommand{\mybignote}[2]{}
\newcommand{\rednote}[2][1=]{}
\newcommand{\bluenote}[2][1=]{}
\newcommand{\greennote}[2][1=]{}
\newcommand{\yellownote}[2][1=]{}

\fi

\usepackage{adjustbox}
\usetikzlibrary{quantikz2}
\usetikzlibrary{backgrounds,fit,decorations.pathreplacing}  
\usetikzlibrary{automata} 
\usetikzlibrary{circuits.logic.CDH}
\usetikzlibrary{circuits.logic.US}
\usetikzlibrary{mindmap}
\usetikzlibrary{decorations}
\usetikzlibrary{decorations.pathmorphing}
\usetikzlibrary{arrows,decorations.pathreplacing}

\tikzset{meter/.append style={draw, inner sep=10, rectangle, font=\vphantom{A}, minimum width=30, line width=.4, path picture={\draw[black] ([shift={(.1,.3)}]path picture bounding box.south west) to[bend left=50] ([shift={(-.1,.3)}]path picture bounding box.south east);\draw[black,-latex] ([shift={(0,.1)}]path picture bounding box.south) -- ([shift={(.3,-.1)}]path picture bounding box.north);}}}

\begin{document}

\title{
High-rate extended binomial codes for multiqubit encoding
} 

\author{En-Jui Chang}
\email{phyenjui@gmail.com}
\affiliation{Independent researcher, Taichung 421786, Taiwan}
 
\date{September 9, 2025}

\begin{abstract}
We introduce a class of bosonic quantum error-correcting codes, termed \emph{extended binomial codes}, which generalize the structure of one-mode binomial codes by incorporating ideas from high-rate qubit stabilizer codes. These codes are constructed in close analogy to $[[n,k,d]]$ qubit codes, where the parameter $n$ corresponds to the total excitation budget rather than the number of physical qubits. Our construction achieves a significant reduction in average excitation per mode while preserving error-correcting capabilities, offering improved compatibility with hardware constraints in the strong-dispersive regime. We demonstrate that extended binomial codes not only reduce the mean excitation required for encoding but also simplify syndrome extraction and logical gate implementation, particularly the logical $\bar{X}$ operation. These advantages suggest that extended binomial codes offer a scalable and resource-efficient approach for bosonic quantum error correction.
\end{abstract}

\maketitle

\section{Introduction}
Designing quantum error-correcting codes (QECCs) for qubits~\cite{Shor95, Ste96, PhysRevA.54.4741, Knill1997} has historically involved collaboration between physicists and classical coding theorists. Two well-known QECCs are the shor code~\cite{Shor95} and the steane code~\cite{Ste96,PhysRevA.54.4741}, both of which use classical error-correcting codes (CECCs) to address bit-flip and phase-flip errors. The shor code achieves this by concatenating two CECCs, one as the inner code and the other as the outer code, while the steane code effectively overlays two CECCs in parallel. These foundational strategies have inspired a wide range of approaches for encoding $k$ logical qubits into $n$ physical qubits, with code distance $d$ guaranteeing the ability to detect up to $\lfloor \frac{d}{2} \rfloor$ Pauli errors and correct up to $\lfloor \frac{d-1}{2} \rfloor$ Pauli errors. Such QECCs are known as $[[ n, k, d ]]$ stabilizer codes.

The collaboration between physicists and classical coding theorists is typically absent in the context of bosonic codes~\cite{Albert2018}. Three of the most well-known bosonic QECCs are the GKP code~\cite{Gottesman2001}, the cat code~\cite{Cochrane1999,Albert2019,PhysRevA.106.062422}, and the binomial code~\cite{Michael2016}. These codes aim to encode a logical qubit within a single bosonic mode (oscillator), treating the use of multiple physical qubits based QECCs as an undesirable overhead. As a result, there is generally less perceived need for input from classical coding theorists. Moreover, the lack of an analog to the \([[ n, k, d ]]\) stabilizer formalism in the bosonic setting has led to the absence of related concepts such as the code rate, i.e., the ratio \(\frac{k}{n}\) of logical to physical qubits. This conceptual gap contributes to limited engagement between the qubit-QECC and bosonic-QECC communities: the former often show less interest in bosonic codes, while the latter tend to focus more on implementing gate operations than on developing new bosonic code constructions.

The current state-of-the-art approach to combining bosonic and qubit QECCs is to use a bosonic code as the inner code and a qubit code as the outer code. In this hybrid scheme, bosonic codes serve as a means to prepare effective ``physical'' qubits with improved error rates, while the outer qubit code provides additional protection. This concatenated structure allows for a trade-off: one may choose a stronger inner bosonic code paired with a weaker outer qubit code, or vice versa. While stronger outer qubit codes typically require greater qubit overhead to achieve higher code distances, stronger inner bosonic codes demand higher mean excitation numbers to enhance information protection. However, increasing the mean excitation number also leads to a higher rate of amplitude-damping (AD) errors~\cite{Chuang1997, Michael2016}, which must be carefully considered in the overall design. The AD rate \(\gamma\) is determined by the relaxation time \(T_1\) and is expressed as \(\gamma=1-e^{-(\Delta t/T_1)}\) for a process lasting \(\Delta t\). This trade-off arises from the fact that higher excitation states couple more strongly to environmental noise.

Scalable quantum communication protocols require encoding an arbitrarily large number of input qubits. To achieve this, one can employ a concatenated scheme using an inner bosonic code and an outer $[[n,k,d]]$ qubit code to encode the input qubits. However, the conceptual gap between bosonic and qubit QECCs has led to an asymmetry: the qubit-QECC community typically focuses on improving the outer $[[n,k,d]]$ code, while the bosonic-QECC community tends to rely on developments in qubit QECCs for advances in outer code design. This division highlights a need for the development of native $[[n,k,d]]$ bosonic QECCs defined directly within the bosonic framework and equipped with the same structural guarantees as their qubit-based counterparts.

In this work, we bridge the gap between the qubit-QECC and bosonic-QECC communities by proposing a new family of bosonic QECCs inspired by our recent high-rate qubit QECCs~\cite{PhysRevA.111.052602}. To improve the code rate of qubit QECCs, we have made significant progress in two directions: the first~\cite{PhysRevA.111.052602} is inspired by the structure of shor codes, while the second~\cite{2503.05249} draws inspiration from steane codes. Notably, we observe that the codewords in the simplest case of our first approach, encoding a single qubit, are closely related to those of the binomial code~\cite{Michael2016}. Table~\ref{table:compare_shor_binomial} compares these codewords. The key insight from our qubit code construction is that sharing a portion of the excitations among multiple encoded logical qubits enables more efficient error correction. Specifically, when correcting weight-$w$ amplitude-damping (AD) errors, sharing a fraction \(\frac{w}{w+1}\) of the total excitations across logical qubits improves the asymptotic code rate by a factor of \(w+1\). This principle can be naturally extended to the design of bosonic codes.

\begin{table*}[ht]
\centering
\caption{Comparison of the smallest one-mode binomial code, the smallest qubit shor code (\([[4,1,2]]\)), and the smallest instance of the extended binomial code proposed in this work. To illustrate the improvement in reducing mean excitation, we also include a comparison with the \([[6,2,2]]\) qubit code alongside the (extended) binomial codes. This comparison illustrates how excitation sharing improves mean excitation cost while preserving error-correcting capabilities.}
\label{table:compare_shor_binomial}
\begin{tabularx}{\textwidth}{ c  c  c  c  c  c  c }
 \hline
 \hline
 QECCs & \hspace{1.3cm} & Logical $\ket{0}$ & \hspace{1.3cm}  & Logical $\ket{1}$ & \hspace{0.3cm} & Mean excitations \\ [1ex]
 \hline
 \textrm{Binomial}~\cite{Michael2016} &   & $\frac{1}{\sqrt{2}}(\ket{0} + \ket{4})$ & & $\ket{2}$ &  & $2$ \\ [1.0ex]
 $[[4,1,2]]$~\cite{PhysRevA.111.052602} &   & $\frac{1}{\sqrt{2}}(\ket{0}^{\otimes 4} + \ket{1}^{\otimes 4})$ & & $\frac{1}{\sqrt{2}}(\ket{0}^{\otimes 2}\ket{1}^{\otimes 2} + \ket{1}^{\otimes 2}\ket{0}^{\otimes 2})$ &  & $2$ \\ [1.0ex]
 (This work) &   & $\frac{1}{\sqrt{2}}(\ket{0}^{\otimes 2} + \ket{2}^{\otimes 2})$ & & $\frac{1}{\sqrt{2}}(\ket{0}\ket{2} + \ket{2}\ket{0})$ &  & $2$\\ [1.0ex]
  &   & Logical $\ket{0}^{\otimes 2}$ & & Logical $\ket{1}^{\otimes 2}$ & &  \\ [1.0ex]
 \hline
 \textrm{Binomial}~\cite{Michael2016} &   & $\frac{1}{\sqrt{2}}(\ket{0} + \ket{4})^{\otimes 2}$ & & $\ket{2}^{\otimes 2}$ &  & $4$ \\ [1.0ex]
 $[[6,2,2]]$~\cite{PhysRevA.111.052602} &   & $\frac{1}{\sqrt{2}}(\ket{0}^{\otimes 6} + \ket{1}^{\otimes 6})$ & & $\frac{1}{\sqrt{2}}(\ket{0}^{\otimes 2}\ket{1}^{\otimes 4} + \ket{1}^{\otimes 2}\ket{0}^{\otimes 4})$ &  & $3$ \\ [1.0ex]
 (This work) &   & $\frac{1}{\sqrt{2}}(\ket{0}^{\otimes 3} + \ket{2}^{\otimes 3})$ & & $\frac{1}{\sqrt{2}}(\ket{0}\ket{2}^{\otimes 2} + \ket{2}\ket{0}^{\otimes 2})$ &  & $3$ \\ [1.0ex]
  &   & Logical $\ket{0}\ket{1}$ &  & Logical $\ket{1}\ket{0}$ &  &  \\ [1.0ex]
 \hline
 \textrm{Binomial}~\cite{Michael2016} &   & $\frac{1}{\sqrt{2}}(\ket{0} + \ket{4})\ket{2}$ & & $\frac{1}{\sqrt{2}}\ket{2}(\ket{0} + \ket{4})$ &  & $4$ \\ [1.0ex]
 $[[6,2,2]]$~\cite{PhysRevA.111.052602} &   & $\frac{1}{\sqrt{2}}(\ket{0}^{\otimes 4}\ket{1}^{\otimes 2} + \ket{1}^{\otimes 4}\ket{0}^{\otimes 2})$ & & $\frac{1}{\sqrt{2}}(\ket{0}^{\otimes 2}\ket{1}^{\otimes 2}\ket{0}^{\otimes 2} + \ket{1}^{\otimes 2}\ket{0}^{\otimes 2}\ket{1}^{\otimes 2})$ &  & $3$ \\ [1.0ex]
 (This work) &   & $\frac{1}{\sqrt{2}}(\ket{0}^{\otimes 2}\ket{2} + \ket{2}^{\otimes 2}\ket{0})$ & & $\frac{1}{\sqrt{2}}(\ket{0}\ket{2}\ket{0} + \ket{2}\ket{0}\ket{2})$ &  & $3$ \\ [1.0ex]
 \hline
 \hline
\end{tabularx}
\end{table*}

Additionally, tracing back the development of the binomial code~\cite{Michael2016}, which uses a single oscillator, reveals that the one-mode binomial code can be viewed as a reduction of the earlier two-mode binomial code~\cite{Chuang1997}. Notably, it took nearly two decades to arrive at a bosonic code that reduces the mean excitation number by half to better mitigate AD errors. However, this reduction comes at a cost: the resulting code loses the ability to protect against collective coherent (CC) errors arising from the system’s intrinsic Hamiltonian. In our previous work on qubit QECCs~\cite{PhysRevA.111.052602}, we encountered a similar trade-off, but in the reverse direction. In the present work, we also construct corresponding bosonic QECCs that deliberately increase the mean excitation number to gain immunity against CC errors, which are otherwise difficult to address in low-excitation codes.

\section{Preliminaries}

\subsection{Composite error model}
In this work, the primary errors we consider are AD errors, arising from unavoidable excitation loss during quantum communication, and CC errors, which stem from the presence of an intrinsic Hamiltonian when the communication time is not assumed to be a fixed, known value (analogous to a classical base station transmitting signals to mobile receivers). We first focus exclusively on AD errors to validate the design of extended binomial codes. We then extend our discussion to CC errors in the context of constant-excitation (CE) codes, which are inherently immune to such errors.

We assume that each oscillator undergoes an independent AD channel, represented by the Kraus operators:
\begin{align}
    \hat{A}_{\ell} = \sum_{k \ge \ell}^{\infty} \sqrt{\binom{k}{\ell}(1-\gamma)^{k-\ell}\gamma^{\ell}} \ket{k-\ell}\bra{k}, \quad k, \ell \in \mathbb{N},
\end{align}
where $\gamma$ is the probability that an excited state $\ket{1}$ relaxes to the ground state $\ket{0}$, and $\mathbb{N}$ denotes the set of natural numbers including $0$.

For an $N$-oscillator system, the error operator $\hat{\mathcal{A}}_{a}$ is defined as:
\begin{align}
    \hat{\mathcal{A}}_{a} = \bigotimes_{j=0}^{N-1}\hat{A}_{a_{j}},
\end{align}
where $a = a_{0}a_{1}\dots a_{N-1} \in \mathbb{N}^{N}$ is a length-$N$ numeral string.

Additionally, intrinsic CC errors are described by the CC channel $\rho \mapsto \hat{U}_{\mathrm{CC}}(\Delta t) \rho \hat{U}_{\mathrm{CC}}^{\dagger}(\Delta t)$, where
\begin{align}
\hat{U}_{\mathrm{CC}}(\Delta t) = \bigotimes_{j=0}^{N-1} e^{-\mathrm{i}\hbar(\hat{n}_{j} + \frac{1}{2}) \Delta t}
\end{align}
is the unitary evolution operator. The number operator $\hat{n}_j$ represents the excitation level of the $j$-th oscillator, with $\Delta t \in \mathbb{R}^{+}$ representing an unknown, positive duration associated with either circuit operation or signal transmission. For simplicity, we set $\hbar = 1$ and omit the global phase factor $e^{-\mathrm{i}\Delta t/2}$, which does not affect the dynamics.

The overall error model assuming CC errors for an $N$-oscillator system is then described by:
\begin{align}
    \mathcal{E}^{\mathrm{CC-AD}}(\rho) = \sum_{a \in \mathbb{N}^{N}} \hat{\mathcal{A}}_{a} \hat{U}_{\mathrm{CC}}(\Delta t)\rho \hat{U}_{\mathrm{CC}}(\Delta t)^{\dagger}\hat{\mathcal{A}}_{a}^{\dagger}.
\end{align}

Let $\wt(a)$ denote the weight of $a$, defined as the sum of all its elements:
\begin{align}
    \wt(a) = \sum_{j=0}^{N-1} a_{j}.
\end{align}

The error operator $\hat{\mathcal{A}}_{a}$ corresponds to a weight-$\wt(a)$ AD error, indicating the loss of $\wt(a)$ excitations.

\subsection{Existing binomial codes encoding a qubit}

Since each excitation in an oscillator can decay to a lower energy state due to coupling with the environment, the coefficient appearing in the Kraus operators for AD errors naturally exhibits a binomial structure. Binomial QECCs leverage this structure by constructing codewords that align with the corresponding statistical distribution. The logical codewords for encoding a single qubit into the one-mode~\cite{Michael2016} and two-mode~\cite{Chuang1997} binomial QECCs are denoted by \(\ket{i}_{\mathrm{1\text{-}bin}}^{(w)}\) and \(\ket{i}_{\mathrm{2\text{-}bin}}^{(w)}\), respectively:
\begin{subequations}
\begin{align}
    \ket{i}_{\mathrm{1\text{-}bin}}^{(w)} &= 
    \sum_{\substack{i \le n \le w+1 \\ (n-i)\,\mathrm{even}}} \sqrt{\frac{\binom{w+1}{n}}{2^w}} \ket{n(w+1)}, \label{eq:1bin} \\[1ex]
    \ket{i}_{\mathrm{2\text{-}bin}}^{(w)} &= 
    \sum_{\substack{i \le n \le w+1 \\ (n-i)\,\mathrm{even}}} \sqrt{\frac{\binom{w+1}{n}}{2^w}} \ket{n(w+1)} \ket{(w+1-n)(w+1)}. \label{eq:2bin}
\end{align}
\end{subequations}
The $(n-i)$ even condition ensures codeword orthogonality.

\subsection{High-rate shor codes}\label{sec:high_rate_shor}

In our previous work on qubit QECCs~\cite{PhysRevA.111.052602}, we also observed a similar binomial structure in the design of the codewords. Specifically, for a weight-\textit{w} AD-correcting code encoding one logical qubit, the logical states are defined as:
\begin{align}\label{eq: qubit_AD_w_1}
    \ket{i}_{\text{AD}}^{(w,1)}
    =& \frac{1}{\sqrt{2^{w}}} \sum_{a \in \{0,1\}^{w} : \atop \wt(a) = 0 \mod 2} \bigotimes_{j=0}^{w-1} \ket{a_j}^{\otimes (w+1)} \ket{i}^{\otimes (w+1)} \notag \\
    &+ \frac{1}{\sqrt{2^{w}}} \sum_{b \in \{0,1\}^{w} : \atop \wt(b) = 1 \mod 2} \bigotimes_{j=0}^{w-1} \ket{b_j}^{\otimes (w+1)} \ket{i'}^{\otimes (w+1)},
\end{align}
where $i' = i \bigoplus 1$ and $i \in \{0,1\}$. To express these codewords in a form more directly analogous to the one-mode binomial code in Eq.\eqref{eq:1bin}, we consider constructing a $(w+1)$-bit string $n$ by appending either $i$ or $i'$ to the $w$-bit strings $a$ or $b$, respectively. This reformulates Eq.\eqref{eq: qubit_AD_w_1} as:
\begin{align}
    \ket{i}_{\text{AD}}^{(w,1)}
    =& \frac{1}{\sqrt{2^{w}}} \sum_{ i \le \wt(n) \le w+1 : \atop \wt(n) = i \mod 2} \bigotimes_{j=0}^{w} \ket{n_j}^{\otimes (w+1)}.
\end{align}

If we reinterpret all $(w+1)^2$ physical qubits as excitations of a single bosonic mode, the resulting state exhibits the same excitation structure as the one-mode binomial code. In particular, each term with Hamming weight $\wt(n) = n$ contributes a component proportional to $\sqrt{\binom{w+1}{n}} \ket{n(w+1)}$, matching the binomial envelope.

Although this example does not yet demonstrate the potential of high-rate codes, it clearly illustrates the structural similarity between these qubit codes and bosonic binomial codes.

Although directly merging all qubits into a single oscillator is not applicable for our high-rate code constructions (a point we elaborate on later), it is still useful to revisit the core structure underlying our previous qubit QECCs.

Suppose an inner code encodes a single qubit into $(w+1)$ qubits using the basis states $\{\ket{0}^{\otimes (w+1)}, \ket{1}^{\otimes (w+1)}\}$. Then, a string of $K$ qubits, indexed by the binary string $i = i_0 i_1 \dots i_{K-1} \in \{0,1\}^K$, is encoded as:
\begin{align}
    \ket{i}_{\text{inner}}^{(w,K)} 
    = \bigotimes_{j=0}^{K-1} \ket{i_j}^{\otimes (w+1)}.
\end{align}

We define the following superposition states:
\begin{align}
    \ket{\pm}_{\text{inner}}^{(w,1)} 
    &= \frac{1}{\sqrt{2}} \left( \ket{0}^{\otimes (w+1)} \pm \ket{1}^{\otimes (w+1)} \right), \notag \\
    \overline{\ket{\pm i}}_{\text{inner}}^{(w,K)} 
    &= \frac{1}{\sqrt{2}} \left( \ket{i}_{\text{inner}}^{(w,K)} \pm \ket{i'}_{\text{inner}}^{(w,K)} \right),
\end{align}
where $i' = i'_0 i'_1 \dots i'_{K-1}$ is the bitwise complement of $i$, i.e., $i'_\ell = i_\ell \oplus 1$ for all $\ell \in \{0, \dots, K-1\}$.

Whereas traditional shor codes use a simple concatenation of an inner repetition code with an outer code to detect and correct both Pauli $X$ and $Z$ errors, our previous work demonstrated that the outer code can be shared among multiple qubits encoded by the inner code. This insight enables a more resource-efficient structure. Specifically, the qubit QECC is spanned by the codewords:
\begin{align}
    \overline{\ket{\pm i}}_{\text{AD}}^{(w,K)} 
    = \left( \ket{\pm}_{\text{inner}}^{(w,1)} \right)^{\otimes w} \otimes \overline{\ket{\pm i}}_{\text{inner}}^{(w,K)}.
\end{align}

Finally, the logical basis states are constructed as:
\begin{align}
    \ket{i}_{\text{AD}}^{(w,K)} 
    = \frac{1}{\sqrt{2}} \left( \overline{\ket{+ i}}_{\text{AD}}^{(w,K)} + (-1)^i\, \overline{\ket{- i}}_{\text{AD}}^{(w,K)} \right).
\end{align}

This construction generalizes the structure presented in Eq.~\eqref{eq: qubit_AD_w_1}. As shown, the qubits are organized into $(w+K)$ groups, each consisting of $(w+1)$ identical copies of a basis state. Consequently, it is more natural to associate each group with a separate oscillator, rather than merging all qubits into a single oscillator as in typical binomial codes.

\section{Extended Binomial QECCs}

Building on the discussion in Sec.~\ref{sec:high_rate_shor}, we now construct the bosonic analog of high-rate qubit QECCs through the following steps:
\begin{enumerate}
    \item[(i)] Define the \emph{inner} code subspace, spanned by the basis states $\{\ket{0}, \ket{w+1}\}$.
    \begin{align}\label{eq:boson_rep}
        \ket{i}_{\text{boson, rep}}^{(w,K)}
        = \bigotimes_{j=0}^{K-1} \ket{i_{j}(w+1)},
    \end{align}
    where $i = i_{0} \dots i_{K-1} \in \{0,1\}^{K}$ is a binary string.
    \item[(ii)] Construct the inner logical states $\ket{\pm}$ that encode multiple qubits, serving as the foundation for the outer code.
    \begin{align}
        \ket{\pm}_{\text{boson, rep}}^{(w,1)} &= \frac{1}{\sqrt{2}} \big( \ket{0} \pm \ket{w+1} \big), \notag\\
        \overline{\ket{\pm i}}_{\text{boson, rep}}^{(w,K)} &= \frac{1}{\sqrt{2}} \big( \ket{i}_{\text{boson, rep}}^{(w,K)} \pm \ket{i'}_{\text{boson, rep}}^{(w,K)} \big),
    \end{align}
    where $i' = i'_{0} \dots i'_{K-1} \in \{0,1\}^{K}$, and $i'_{\ell} = i_{\ell} + 1 \mod 2$ for all $\ell \in \{0, \dots, K-1\}$.
    \item[(iii)] Define the full extended binomial codewords and express them explicitly in the Fock (number) basis.
    \begin{align}\label{eq: extended_w_K}
        \overline{\ket{\pm i}}_{\text{ext.-bin}}^{(w,K)} &= \big( \ket{\pm}_{\text{boson, rep}}^{(w,1)} \big)^{\otimes w} \otimes \overline{\ket{\pm i}}_{\text{boson, rep}}^{(w,K)},\notag\\
        \ket{i}_{\text{ext.-bin}}^{(w,K)} &= \frac{1}{\sqrt{2}} \big( \overline{\ket{+ i}}_{\text{ext.-bin}}^{(w,K)} + (-1)^{i}\overline{\ket{- i}}_{\text{ext.-bin}}^{(w,K)} \big).
    \end{align}
\end{enumerate}

\subsection{Approximate quantum error-correction conditions}
To assess the effectiveness of the proposed code construction, we verify the approximate quantum error-correction (AQEC) conditions. Theorem~\ref{Thm: AQEC} establishes that the extended binomial code approximately corrects any error in the set of AD errors of weight up to $w$, by satisfying the Knill-Laflamme conditions~\cite{Knill1997, Got97}. The detailed proof is provided in the Appendix.

\begin{theorem}\label{Thm: AQEC}
\textrm{AQEC conditions for the AD $w$-code for qudits or oscillators.}
Let $\big\{\ket{i}_{\mathrm{ext.-bin}}^{(w,K)}| i \in\{0,1\}^{K}\big\}$ be its codewords. Define the correctable error set $\mathcal{A} = \big\{\hat{\mathcal{A}}_{a} : a \in \mathbb{N}^{(w+K)}, \wt(a) \leq w \big\}$. Then, for any $\hat{\mathcal{A}}_{k}, \hat{\mathcal{A}}_{\ell}\in\mathcal{A}$,  we have
\begin{align}
{\bra{i}}_{\mathrm{ext.-bin}}^{(w,K)}\hat{\mathcal{A}}_{k}^{\dagger}\hat{\mathcal{A}}_{\ell}{\ket{j}}_{\mathrm{ext.-bin}}^{(w,K)}
=&\delta_{ij}C_{k\ell}+O(\gamma^{w+1}),
\end{align}
where $\delta_{ij}$ is the Kronecker delta function, and $C_{k\ell}$ is a complex number independent of these two length-$K$ natural-number strings $i$ and $j$.
\end{theorem}

\subsection{Stabilizers and syndrome extraction}

While the codewords are adapted from our previous qubit QECCs~\cite{PhysRevA.111.052602}, the stabilizer structure is simplified in the present construction. As discussed earlier, the code uses $(w+K)$ oscillators, each corresponding to one of the $(w+K)$ qubit groups.

The key stabilizers are those that span multiple groups (or oscillators). For $w \ge 2$, the stabilizer group includes $w$ generators given by
\begin{subequations}
\begin{align}
    \hat{S}_{i} &= \hat{\textbf{X}}_{i}\hat{\textbf{X}}_{i+1}, \quad i \in \{0, \dots, w-2\},\\
    \hat{S}_{w-1} &= \bigotimes_{i=w-1}^{w-1+K} \hat{\textbf{X}}_{i},
\end{align}
\end{subequations}
where $\hat{\textbf{X}} = \left(\ket{w+1}\bra{0} + \ket{0}\bra{w+1} + \hat{H}_{\perp}\right)$ is an analog of the Pauli-$X$ operator, and $\hat{H}_{\perp}$ is an arbitrary Hermitian operator orthogonal to the code space.

These stabilizers enforce excitation parity constraints across neighboring oscillators. Specifically, for the first $w$ oscillators, $\hat{S}_i$ implies that the excitation difference between adjacent oscillators must be $\pm(w+1)$. This can be diagnosed by measuring the set of observables:
\begin{align}
    \left\{ \hat{O}_{i} = \left(\hat{n}_{i} - \hat{n}_{i+1}\right)^2 \bmod (w+1) \;\middle|\; i \in \{0, \dots, w-2\} \right\},
\end{align}
which detect relative excitation loss events. These observables involve combinations of self-Kerr ($\hat{n}_i^2$) and cross-Kerr ($\hat{n}_i \hat{n}_{i+1}$) interactions.

From a reduction-based perspective in theoretical computer science, if one can measure both $n_i \bmod (w+1)$ and $n_{i+1} \bmod (w+1)$ individually, then one can compute $(n_i - n_{i+1})^2 \bmod (w+1)$. However, the converse does not hold. In this sense, measuring $(n_i - n_{i+1})^2 \bmod (w+1)$ extracts less detailed information and is therefore considered a simpler measurement in terms of informational content.

From the experimental perspective, the notion of measurement difficulty is different. We expect that the challenge of measuring cross-Kerr-type observables, such as joint photon-number parities or excitation differences, is comparable to other demonstrated tasks. Examples include preparing two-mode binomial codewords (as discussed in Sec.~IX of Ref.~\cite{Chuang1997}) and implementing Kerr-type controlled-phase gates (see Table~I in Ref.~\cite{Albert2019}). While a detailed hardware-specific analysis is beyond the scope of this theoretical work, we believe that these measurements are achievable with a level of difficulty comparable to that of those required for other bosonic QECCs.

The final stabilizer $\hat{S}_{w-1}$ enforces a global parity constraint between the $(w-1)$th oscillator and the set of $K$ subsequent oscillators. For $K=1$, the allowed excitation difference is in $\{0, \pm(w+1)\}$; for general $K \ge 2$, it extends to $\{0, \pm(w+1), \dots, \pm K(w+1)\}$. The corresponding observable is
\begin{align}
    \hat{O}_{w-1} = \left( \hat{n}_{w-1} - \sum_{i=w}^{w-1+K} \hat{n}_{i} \right)^2 \bmod (w+1),
\end{align}
which also indicates relative loss events.

These stabilizer measurements can be implemented using bosonic cnot gates and ancillary qubits, as demonstrated in the two-mode binomial code~\cite{Chuang1997}. An AD error is detected when one or more stabilizer values deviate from their expected eigenvalues. The corresponding observable outcomes identify AD events up to weight $w$.

In the first $w$ oscillators, AD events can be located using a classical lookup table. For the remaining $K$ oscillators, AD events are detected but not precisely localized, since the last stabilizer $\hat{S}_{w-1}$ only signals the presence of errors within this group without resolving their exact positions. After correcting the first $w$ oscillators based on stabilizer outcomes, partial decoding of the outer repetition code isolates the last $K$ oscillators. AD errors within these data oscillators can then be identified using the inner code structure defined in Eq.~\eqref{eq:boson_rep}. Conditional excitations are subsequently applied to complete the correction process and recover the encoded quantum information.

We can compare the syndrome extraction procedure of our construction with that of the original one-mode binomial codes when encoding $K$ logical qubits. In both cases, measuring excitation number modulo $(w+1)$ is a shared feature. However, our code requires measuring excitation differences between neighboring oscillators, yielding $w$ excitation checks, whereas the original one-mode binomial codes require $K$ individual excitation measurements on each of the $K$ oscillators.

The trade-off is as follows: our code uses $(w+K)$ oscillators instead of only $K$, increasing the hardware overhead. On the other hand, each oscillator in the original one-mode binomial code must support up to $(w+1)^2$ excitations, while in our extended binomial code, each oscillator only needs to support up to $(w+1)$ excitations. This lower excitation requirement may enable the use of more experimentally feasible excitation measurements, particularly in regimes where measurement fidelity degrades at high excitation levels. 

Suppose that each excitation check may be prone to error depending on the maximum excitation number per oscillator. In Table~\ref{table:compare_syndrome}, we summarize the trade-offs between the syndrome extraction strategies used in the original one-mode binomial codes and the extended binomial codes proposed in this work.

\begin{table}[ht]
\caption{Comparison of syndrome extraction strategies between the original one-mode binomial code and the extended binomial code proposed in this work. While the two approaches differ in excitation measurement complexity, precise performance metrics require experimental benchmarking.}
\label{table:compare_syndrome}
\begin{tabularx}{\linewidth}{ l c c c }
 \hline
 \hline
  & \hspace{0.5cm} & \multicolumn{2}{c}{Binomial}\\ [1.0ex]
 Property & & Ref.~\cite{Michael2016} & This work \\ [1.0ex]
 \hline
 &&&\\
 Maximum excitation per oscillator & & $(w+1)^2$ & $w+1$ \\ [1.0ex]
 Measurement difficulty & & Higher & Lower \\ [1.0ex]
 Oscillator requiring measurement & & $K$ & $K+w$ \\ [1.0ex]
 \hline
 \hline
\end{tabularx}
\end{table}

\subsubsection*{Measurement difficulty of excitations}

To clarify the significance of excitation measurement complexity, we revisit a key assumption made in the original one-mode binomial code proposal~\cite{Michael2016}. That proposal assumes the availability of high-fidelity quantum nondemolition measurements of photon number parity, facilitated by conditional unitary control of the binomial code. This capability is attributed to the strong dispersive coupling between an ancillary qubit and the bosonic mode, described by the Hamiltonian $\hat{H}_{\mathrm{disp}}/\hbar = \chi\hat{Z}\otimes\hat{n}$, where $\chi=\frac{g^2}{\Delta}$.

While this assumption is valid in the strong-dispersive regime, it is natural to ask: how strong must this coupling be, and under what conditions does the assumption break down? Reference~\cite{PhysRevA.74.042318}, cited in Ref.~\cite{Michael2016}, provides a theoretical foundation using the number splitting technique. There, the dispersive Hamiltonian is derived by approximately diagonalizing the full Jaynes-Cummings model to second order in the small parameter $\frac{g}{\Delta}$, where $g$ is the coupling strength and $\Delta$ is the detuning between the qubit and the cavity mode. As $g$ increases or $\Delta$ decreases, the validity of the dispersive approximation degrades and photon loss becomes more likely.

To quantify this limitation, a critical excitation number is defined as $n_{\mathrm{c}} = \frac{\Delta^2}{4g^2}$, which upper bounds the average photon number in the cavity consistent with the dispersive regime. Based on experimental results, Ref.~\cite{PhysRevA.74.042318} estimates $n_{\mathrm{c}} \approx 82$. For the one-mode binomial code, which has an average excitation number $\langle n \rangle \sim \frac{(w+1)^2}{2}$, this constrains the maximum correctable error weight to approximately $w_{\mathrm{1\text{-}bin}} = \lfloor \sqrt{2 n_{\mathrm{c}}} \rfloor - 1 = 11$.

In contrast, our extended binomial codes distribute the excitation across multiple modes. For the same error weight $w$, the average and peak excitation per mode scale only as $O(w)$. This allows a significantly higher excitation budget while remaining within the dispersive limit. Specifically, assuming each mode must satisfy the same $n_{\mathrm{c}}$ bound, the total allowed excitation can reach up to $2n_{\mathrm{c}}$, permitting $w_{\mathrm{ext.-bin}} = \lfloor 2n_{\mathrm{c}} \rfloor - 1 = 163$.

This dramatic increase in error-correcting capacity, from $w_{\mathrm{1\text{-}bin}} = 11$ to $w_{\mathrm{ext.-bin}} = 163$, comes with the trade-off that excitation parity must be measured across $K + w$ modes, rather than just $K$ as in the one-mode case. However, when $K \gg w$, this overhead becomes negligible, making extended binomial codes highly advantageous in regimes constrained by dispersive readout fidelity.

\subsection{Logical operators}
For the $K$ logical qubits, we denote the logical Pauli operators by $\bar{X}_{\ell}$ and $\bar{Z}_{\ell}$ for $\ell \in \{0, \dots, K-1\}$, with $\bar{Y}_{\ell} = -\mathrm{i}\bar{Z}_{\ell}\bar{X}_{\ell}$. The logical Hadamard operator is defined as $\bar{H}_{\ell} = \frac{1}{\sqrt{2}}(\bar{X}_{\ell} + \bar{Z}_{\ell})$. These logical $\bar{X}$ and $\bar{Z}$ operations can be implemented using $(w+1)$-local operators:
\begin{subequations}
\begin{align}
    \bar{X}_{\ell} &= \hat{\textbf{X}}_{w+\ell}, \\
    \bar{Z}_{\ell} &= \bigotimes_{j=0}^{w-1} \hat{\textbf{Z}}_{j} \otimes \hat{\textbf{Z}}_{w+\ell},
\end{align}
\end{subequations}
where $\hat{\textbf{Z}} = \left(\ket{0}\bra{0} - \ket{w+1}\bra{w+1} + \hat{H}_{\perp}\right)$ acts as an analog of the Pauli-$Z$ operator for a qubit, and can be implemented as $e^{\mathrm{i}(\frac{\pi}{w+1})\hat{n}}$. The operator $\hat{\textbf{X}}$ flips the basis states as $\ket{0} \leftrightarrow \ket{w+1}$.

Notably, this implementation of the logical $\bar{X}$ operator differs from that in the original binomial code, which requires a more intricate local transformation, such as $\frac{1}{\sqrt{2}}(\ket{0} + \ket{4}) \leftrightarrow \ket{2}$, within a single oscillator. This difference becomes increasingly significant as the parameter $w$ grows larger.

Additionally, the global logical operator $\bar{X}_{\text{all}} = \bigotimes_{\ell=0}^{K-1} \bar{X}_{\ell}$ can be implemented efficiently due to the $X$-type stabilizers:
\begin{align}
    \bar{X}_{\mathrm{all}} &= \hat{\textbf{X}}_{0}.
\end{align}
This efficient global operator $\bar{X}_{\mathrm{all}}$ requires only a single local $\hat{\textbf{X}}$ operator, instead of $K$ separate $\hat{\textbf{X}}$ operators. When $K$ is much larger than $1$, this global operator $\bar{X}_{\mathrm{all}}$ is significantly more efficient than flipping each of the $K$ qubits individually, assuming the encoding and decoding costs are negligible.

We summarize the two approaches for implementing logical gates in the original binomial codes and our extended binomial codes in Table~\ref{table:compare_logical_operator}. Notably, the implementation of the logical $\bar{X}$ gate in our construction is significantly simpler than in the one-mode binomial codes, where the logical states [e.g., Eq.\eqref{eq:1bin}] involve nontrivial superpositions. As noted in the original binomial code proposal\cite{Michael2016}, ``The difficulty of achieving such unitaries is on the same scale as the gates needed to perform the encoding of the initial state," indicating that implementing a logical $\bar{X}$ gate requires a nontrivial unitary transformation. In contrast, our scheme requires only a basis-state flip between $\ket{0}$ and $\ket{w+1}$, resulting in a far simpler implementation. This gap in complexity becomes even more significant as either $w$ or $K$ increases, especially for global operations such as $\bar{X}_{\mathrm{all}}$.

\begin{table}[ht]
\caption{Comparison of strategies for implementing logical $\bar{X}$ and $\bar{Z}$ gates in the original one-mode binomial code and the extended binomial code proposed in this work. The implementation cost of $\bar{X}_{\mathrm{all}}$ grows substantially in the original binomial code as $w$ or $K$ increases, whereas it remains simple in our construction.}
\label{table:compare_logical_operator}
\begin{tabularx}{\linewidth}{ c c c c }
 \hline
 \hline
 & \hspace{0.5cm} & \multicolumn{2}{c}{Operator}\\ [1.0ex]
 & & $\bar{X}_{\ell}$ & $\bar{Z}_{\ell}$ \\ [1.0ex]
 \hline\\
 Ref.~\cite{Michael2016} & & $(\ket{0}_{\mathrm{1\text{-}bin}}^{(w)}\leftrightarrow\ket{1}_{\mathrm{1\text{-}bin}}^{(w)})_{\ell}$ & $e^{\mathrm{i}(\frac{\pi}{w+1})\hat{n}_{\ell}}$ \\  [1.0ex]
 This work & & $(\ket{0}\leftrightarrow\ket{w+1})_{w+\ell}$ & $e^{\mathrm{i}(\frac{\pi}{w+1})\left(\sum_{j=0}^{w-1}\hat{n}_{j}+\hat{n}_{w+\ell}\right)}$ \\ [1.0ex]
 & & \multicolumn{2}{c}{Operator}\\ [1.0ex]
 & & $\bar{X}_{\mathrm{all}}$ & $\bar{Z}_{\mathrm{all}}$ \\ [1.0ex]
 \hline\\
 Ref.~\cite{Michael2016} & & $(\ket{0}_{\mathrm{1\text{-}bin}}^{(w)}\leftrightarrow\ket{1}_{\mathrm{1\text{-}bin}}^{(w)})^{\otimes K}$ & $e^{\mathrm{i}(\frac{\pi}{w+1})\left(\sum_{\ell=0}^{K-1}\hat{n}_{\ell}\right)}$ \\ [1.0ex]
 This work & & $(\ket{0}\leftrightarrow\ket{w+1})_{0}$ & $\bigotimes_{\ell=0}^{K-1}\bar{Z}_{\ell}$ \\ [1.0ex]
 \hline
 \hline
\end{tabularx}
\end{table}

\subsection{Encoding circuit}

Having established the relative complexities of measurement and logical operators, we now analyze the complexity of preparing a logical qubit $\ket{\bar{\psi}}$ from an unknown physical qubit $\ket{\psi}$. The goal is to provide a clear comparison of the challenges involved in encoding a logical qubit using the standard binomial code and our extended construction.

As a baseline, we propose a simple, abstract encoding scheme for a single qubit in both cases. Notably, the original proposal of the binomial code~\cite{Michael2016} did not specify an explicit encoding circuit or a state-preparation strategy in the presence of faulty gates (Fig.~3(b) of Ref.~\cite{Michael2016} simply shows a downward arrow labeled \textit{Encoding}, without further elaboration). Our proposed procedure proceeds as follows.
\begin{enumerate}
\item[(i)] Start with a physical qubit $\ket{\psi}=\alpha\ket{0}+\beta\ket{1}$ to be encoded, together with a logical qubit $\ket{\bar{+}}$ prepared from postselection by measuring logical $\bar{X}$. The initial joint state is $\frac{1}{\sqrt{2}}(\alpha\ket{0}+\beta\ket{1})\otimes(\ket{\bar{0}}+\ket{\bar{1}})$.
\item[(ii)] Measure $\hat{Z}\otimes\bar{Z}$ in preparation for the controlled logical gate. The post-measurement states are
\begin{align*}
\tfrac{1}{\sqrt{2}}(\alpha\ket{0}\ket{\bar{0}}+\beta\ket{1}\ket{\bar{1}}),\quad &\text{outcome: }+1,\\
\tfrac{1}{\sqrt{2}}(\alpha\ket{0}\ket{\bar{1}}+\beta\ket{1}\ket{\bar{0}}),\quad &\text{outcome: }-1.
\end{align*}
Since the physical-qubit component is fixed across codewords, the relevant complexity lies in measuring logical $\bar{Z}$.
\item[(iii)] If the outcome is $+1$, do nothing; if it is $-1$, apply a logical $\bar{X}$. The state then becomes
\begin{equation*}
\tfrac{1}{\sqrt{2}}(\alpha\ket{0}\ket{\bar{0}}+\beta\ket{1}\ket{\bar{1}}).
\end{equation*}
The relevant complexity here is that of implementing logical $\bar{X}$.
\item[(iv)] Measure $\hat{X}$ on the physical qubit and discard it. The post-measurement states are
\begin{align*}
\tfrac{1}{\sqrt{2}}(\alpha\ket{\bar{0}}+\beta\ket{\bar{1}}),\quad &\text{outcome: }+1,\\
\tfrac{1}{\sqrt{2}}(\alpha\ket{\bar{0}}-\beta\ket{\bar{1}}),\quad &\text{outcome: }-1.
\end{align*}
This step does not depend on the choice of code and is thus not relevant for the comparison.
\item[(v)] If the outcome is $+1$, do nothing; if it is $-1$, apply a logical $\bar{Z}$. The final state is the desired logical qubit $\ket{\bar{\psi}}=\alpha\ket{\bar{0}}+\beta\ket{\bar{1}}$. The relevant complexity here is that of implementing logical $\bar{Z}$.
\end{enumerate}

From this procedure, it follows that the difficulty of encoding reduces to three factors: the complexity of measuring logical $\bar{Z}$, and of implementing logical $\bar{X}$ and $\bar{Z}$. Based on our earlier comparison of these operations for the binomial and extended binomial codes (see Table~\ref{table:compare_logical_operator}), it is evident that logical $\bar{X}$ is significantly simpler to implement in the extended binomial code. The application of logical $\bar{Z}$, however, is comparable in both cases, as it corresponds to leaving the modes non-interacting under a Hamiltonian proportional to $\hat{n}$.

Therefore, we conclude that state preparation is expected to be easier in the extended binomial code, provided the reduced complexity of logical $\bar{X}$ is indeed a dominant factor.

Note that the above encoding scheme applies to any code with well-defined logical $\bar{X}$ and $\bar{Z}$ operators, although more efficient encoding strategies may exist.

\section{Constant-excitation extended binomial codes}

Having introduced the core ideas behind the extended binomial codes, we now present their CE counterparts. These CE codes follow a construction philosophy similar to that of the transition from one-mode binomial codes in Eq.~\eqref{eq:1bin} to two-mode binomial codes in Eq.~\eqref{eq:2bin}. They can also be viewed as arising from grouping qubits in the CE qubit codes developed in our earlier work. The CE construction is particularly advantageous in scenarios where CC errors are more dominant than AD errors, such as in communication protocols where energy relaxation is already well suppressed at the physical layer.

To explicitly describe the CE counterparts of the extended binomial codewords in Eq.~\eqref{eq: extended_w_K}, we first express those codewords in the number basis:
\begin{align}
&\sqrt{2^{w+2}} \ket{i}_{\mathrm{AD}}^{(w,K)} \notag \\
=&\Bigg( \sum_{a_0,\dots,a_{w-1} \in \{0,1\} \atop \wt(a) \equiv \wt(i) \bmod 2}\bigotimes_{j=0}^{w-1} \ket{a_j(w+1)} \Bigg)\bigotimes_{j=0}^{K-1} \ket{i_j(w+1)} \notag \\
&+ \left( \sum_{a_0,\dots,a_{w-1} \in \{0,1\} \atop \wt(a) \equiv \wt(i)+1 \bmod 2} \bigotimes_{j=0}^{w-1} \ket{a_j(w+1)} \right)
\bigotimes_{j=0}^{K-1} \ket{i'_j(w+1)},
\end{align}
where $i = i_0 \dots i_{K-1} \in \{0,1\}^K$ is a binary string, $i'$ is its bitwise complement, and similarly $a' = a'_0 a'_1 \dots a'_{w-1}$ denotes the bitwise complement of $a$.

To obtain the CE counterpart, we apply a extension by pairing each oscillator with its complement, effectively mapping $\left(\ket{0}, \ket{w+1}\right) \mapsto \left(\ket{0}\ket{w+1}, \ket{w+1}\ket{0}\right)$. This results in the CE codewords:
\begin{align}
&\sqrt{2^{w+2}} \ket{i}_{\mathrm{CE\text{-}AD}}^{(w,K)} \notag \\
=&\Bigg( \sum_{a_0,\dots,a_{w-1} \in \{0,1\} \atop \wt(a) \equiv \wt(i) \bmod 2}\bigotimes_{j=0}^{w-1} \ket{a_j(w+1)}\ket{a'_j(w+1)} \Bigg)\notag\\
&\bigotimes_{j=0}^{K-1} \ket{i_j(w+1)}\ket{i'_j(w+1)} \notag \\
&+ \left( \sum_{a_0,\dots,a_{w-1} \in \{0,1\} \atop \wt(a) \equiv \wt(i)+1 \bmod 2} \bigotimes_{j=0}^{w-1} \ket{a_j(w+1)}\ket{a'_j(w+1)} \right)\notag\\
&\bigotimes_{j=0}^{K-1} \ket{i'_j(w+1)}\ket{i_j(w+1)}.
\end{align}

This construction maintains constant total excitation across all modes and is well-suited for avoiding CC errors.

\section{Extended binomial codes encoding a qudit}

While we refer to our bosonic codes that encode qubits as \emph{extended binomial codes}, this should not be confused with the \emph{extended binomial codes encoding a qudit} introduced in Appendix C of the one-mode binomial code proposal~\cite{Michael2016}. In their main text, the authors consider encoding $k$ qubits into a single resonator using $2^k$ Fock states, spanning photon numbers from $0$ to $2^k - 1$. While the original one-mode binomial codes require a spacing of $(w+1)$ between logical levels to maintain orthogonality and correct errors, the qudit encoding demands $(2^k - 1)$ such spacings to accommodate all $2^k$ logical levels. This results in a substantially higher mean excitation, which exacerbates physical limitations such as the breakdown of the strong-dispersive regime.

In contrast, our extended binomial codes achieve comparable or superior code rates while maintaining a significantly lower average excitation per mode. This makes our construction more robust under realistic experimental constraints. Despite the similarity in terminology, our approach offers a solution that is more resource-efficient and scalable than the one-mode binomial encoding of a qudit.

\section{Conclusion}

We have proposed a family of bosonic quantum error-correcting codes, which we call \emph{extended binomial codes}. These codes are motivated by structural similarities between bosonic codes and qubit-based shor codes. Building on insights from our previous work on high-rate shor codes, we identified a more efficient approach to constructing bosonic QECCs by leveraging techniques from qubit stabilizer codes. This represents an explicit example of bosonic codes constructed in a form analogous to $[[n,k,d]]$ qubit stabilizer codes.

In both qubit and bosonic QECCs, the parameters $k$ (number of logical qubits) and $d$ (code distance) are standard. However, the interpretation of $n$ (resource cost) differs: in qubit codes, $n$ counts the number of physical qubits, while in bosonic codes, it reflects the excitation number budget. With this correspondence in mind, our extended binomial codes achieve significant improvements in excitation efficiency.

To highlight the reduction in average excitation number, we compared representative codewords in Table~\ref{table:compare_shor_binomial}. This efficiency directly translates to advantages in syndrome measurement, as shown in Table~\ref{table:compare_syndrome}. Additionally, our construction simplifies the implementation of certain logical operators, particularly the logical $\bar{X}$ gate, which is notoriously complex in conventional binomial codes. As demonstrated in Table~\ref{table:compare_logical_operator}, the extended binomial codes substantially reduce this complexity.

Overall, our approach offers a scalable and hardware-conscious framework for bosonic QEC, with notable benefits in resource efficiency, syndrome extraction, and logical gate implementation.

\section*{Data Availability}
No data were created or analyzed in this study.



\appendix
\numberwithin{equation}{section}

\section{Approximate quantum error-correction conditions for the bosonic AD $w$-code}\label{app:AQEC}
\textit{Theorem 1.}
\textrm{AQEC conditions for the bosonic AD $w$-code.}
Let $\big\{\ket{i}_{\mathrm{ext.-bin}}^{(w,K)}| i \in\{0,1\}^{K}\big\}$ be its codewords. Define the correctable error set $\mathcal{A} = \big\{\hat{\mathcal{A}}_{a} : a \in \mathbb{N}^{(w+K)}, \wt(a) \leq w \big\}$. Then, for any $\hat{\mathcal{A}}_{k}, \hat{\mathcal{A}}_{\ell}\in\mathcal{A}$,  we have
\begin{align}
{\bra{i}}_{\mathrm{ext.-bin}}^{(w,K)}\hat{\mathcal{A}}_{k}^{\dagger}\hat{\mathcal{A}}_{\ell}{\ket{j}}_{\mathrm{ext.-bin}}^{(w,K)}
=&\delta_{ij}C_{k\ell}+O(\gamma^{w+1}),
\end{align}
where $\delta_{ij}$ is the Kronecker delta function, and $C_{k\ell}$ is a complex number independent of these two length-$K$ natural-number strings $i$ and $j$.

\textbf{Proof.}
Similar to we have done in our previous high-rate qubit code, our strategy for proving that the AQEC conditions are satisfied proceeds in three steps:
\begin{itemize}
    \item[(i)] Verify the case $i\neq j$.
    \item[(ii)] Verify the case $i= j$ and $k\neq\ell$.
    \item[(iii)] Verify the case $i= j$ and $k=\ell$.
\end{itemize}
We begin with the first step, which is straightforward. One can verify that 
\[
{\bra{i}}_{\mathrm{ext.-bin}}^{(w,K)}\hat{\mathcal{A}}_{k}^{\dagger}\hat{\mathcal{A}}_{\ell}{\ket{j}_{\mathrm{ext.-bin}}^{(w,K)}}=0
\]
for $i\neq j$.
This is because $\hat{\mathcal{A}}_{k}{\ket{i}}_{\mathrm{ext.-bin}}^{(w,K)}$ represents the collapse of ${\ket{i}}_{\mathrm{ext.-bin}}^{(w,K)}$ into a superposition of basis states of the form $\bigotimes_{j=0}^{w+K-1} \widetilde{\ket{a_j(w+1)}}$, where each $\widetilde{\ket{a_j(w+1)}}$ is either $\ket{0}$ or a number basis vector of weight at least $1$, indicating $\ket{a_j(w+1)}=\ket{w+1}$. This allows us to identify the original logical state. Thus, the inner product vanishes, as required.

The second step proceeds similarly. We again find that the inner product vanishes,
\[
{\bra{i}}_{\mathrm{ext.-bin}}^{(w,K)}\hat{\mathcal{A}}_{k}^{\dagger}\hat{\mathcal{A}}_{\ell}{\ket{i}}_{\mathrm{ext.-bin}}^{(w,K)}=0
\]
because two length-$(w+K)$ natural-number strings are not equal, i.e. $k\neq \ell$. This orthogonality arises from the distinct Kraus operators $\hat{\mathcal{A}}_{k}$ and $\hat{\mathcal{A}}_{\ell}$, yielding orthogonal error syndromes in the code space. As a result, the inner product between the resulting error states remains $0$.

Finally, the third step is somewhat more involved. Let us define ${\bra{0}}_{\mathrm{ext.-bin}}^{(w,K)}\hat{\mathcal{A}}_{k}^{\dagger}\hat{\mathcal{A}}_{k}{\ket{0}}_{\mathrm{ext.-bin}}^{(w,K)}=C_{kk}$ for a complex number $C_{kk}$. It remains to show that
\begin{align}
    {\bra{i}}_{\mathrm{ext.-bin}}^{(w,K)}\hat{\mathcal{A}}_{k}^{\dagger} \hat{\mathcal{A}}_{k}{\ket{i}}_{\mathrm{ext.-bin}}^{(w,K)}=&C_{kk}+O(\gamma^{w+1}),
\end{align}
where $i \neq 0$ and the operator $\hat{\mathcal{A}}_{k}^{\dagger}\hat{\mathcal{A}}_{k}$ takes the tensor-product form
\begin{align}
   \hat{\mathcal{A}}_{k}^{\dagger}\hat{\mathcal{A}}_{k}=\bigotimes_{j=0}^{w+K-1}\Big[ \sum_{n\ge k_{j}}^{\infty}\binom{n}{k_{j}}(1-\gamma)^{n-k_{j}}\gamma^{k_{j}}\ket{n}\bra{n})\Big].\label{eq:AkAK}
\end{align}

For the logical zero state, we have
\begin{align}
    C_{kk}
    =&\frac{1}{ 2^{w}}  \sum_{a_{0},\dots,a_{w-1}\in\{0,1\};\; a_{w}\in\{0,2^{K}-1\}:\atop \wt(a)=0\mod 2 }\prod_{j=0}^{w} \alpha(a_j),
\end{align}
where
\begin{align}
    \alpha(a_j)
    =& \bra{a_j(w+1)}\left(\hat{\mathcal{A}}_{k_{j}}^{\dagger} \hat{\mathcal{A}}_{k_{j}}\right)\ket{a_j(w+1)},\notag\\
    \forall j \in&\{0,\dots,w-1\},
\end{align}
\begin{align}
    \alpha(a_{w})&\notag\\
    =& \prod_{j=0}^{K-1}\bra{(a_{w})_{j}(w+1)}\left(\hat{\mathcal{A}}_{k_{w+j}}^{\dagger} \hat{\mathcal{A}}_{k_{w+j}}\right)\ket{(a_{w})_{j}(w+1)},
\end{align}
and $a_{w}$ is a length-$K$ natural number string.

Similarly, for a general logical $0$ or $1$ state,
\begin{align}
    &{\bra{i}}_{\mathrm{ext.-bin}}^{(w,K)}\hat{\mathcal{A}}_{k}^{\dagger}\hat{\mathcal{A}}_{k}{\ket{i}}_{\mathrm{ext.-bin}}^{(w,K)}\notag\\
    =&\frac{1}{{2^{w}}}\sum_{b_{0},\dots,b_{w-1}\in\{0,1\};\;b_{w}\in\{i,i'\}:\atop \wt(b)=0\mod 2 }\prod_{j=0}^{w} \alpha(b_j),
\end{align}
where $i'=i+1 \mod 2^{K}$, $\forall j \in\{0,\dots,w-1\}$ and 
\begin{align}
    \alpha(b_j)
    =& \bra{b_j(w+1)}\left(\hat{\mathcal{A}}_{k_{j}}^{\dagger} \hat{\mathcal{A}}_{k_{j}}\right)\ket{b_j (w+1)},
\end{align}
\begin{align}
    \alpha(b_{w})
    =&\prod_{j=0}^{K-1}\bra{(b_{w})_{j} (w+1)}\left( \hat{\mathcal{A}}_{k_{w+j}}^{\dagger} \hat{\mathcal{A}}_{k_{w+j}}\right)\ket{(b_{w})_{j} (w+1)}.
\end{align}

Taking the difference, we obtain
\begin{align*}&2^{w}\left(\overline{\bra{0}}\hat{\mathcal{A}}_{k}^{\dagger}\hat{\mathcal{A}}_{k}\overline{\ket{0}}- \overline{\bra{i}}\hat{\mathcal{A}}_{k}^{\dagger}\hat{\mathcal{A}}_{k}\overline{\ket{i}}\right) \notag\\
=& \sum_{a_{0},\dots,a_{w-1}\in\{0,1\}:\atop \wt(a)=0\mod 2 }\prod_{j=0}^{w-1} \alpha(a_j)\left[\alpha(a_{w}=0)-\alpha(b_{w}=i)\right]\notag\\
+&\sum_{b_{0},\dots,b_{w-1}\in\{0,1\}:\atop \wt(b)=1\mod 2 }\prod_{j=0}^{w-1} \alpha(b_j)\left[\alpha(a_{w}=2^{K}-1)-\alpha(b_{w}=i')\right].
\end{align*}
It can be shown that each difference term, such as $\left[\alpha(a_{w}=0)-\alpha(b_{w}=i)\right]$ for $i\neq 0$, is of the order of $O(\gamma)$.

Consequently,
\begin{align*}
&2^{w}\left(\overline{\bra{0}}\hat{\mathcal{A}}_{k}^{\dagger}\hat{\mathcal{A}}_{k}\overline{\ket{0}}- \overline{\bra{i}}\hat{\mathcal{A}}_{k}^{\dagger}\hat{\mathcal{A}}_{k}\overline{\ket{i}}\right)\notag\\
=& O(\gamma) \sum_{a\in\{0,1\}^{w}:\atop \wt(a)=0\mod 2 }\prod_{j=0}^{w-1} \alpha(a_j)
- O(\gamma) \sum_{b\in\{0,1\}^{w}:\atop \wt(b)=1\mod 2 }\prod_{j=0}^{w-1} \alpha(b_j).
\end{align*}

By iterating this argument, one finds that the total difference is $O(\gamma^{w+1})$, i.e.,
$2^{w}\left(\overline{\bra{0}}\hat{\mathcal{A}}_{k}^{\dagger}\hat{\mathcal{A}}_{k}\overline{\ket{0}}- \overline{\bra{i}}\hat{\mathcal{A}}_{k}^{\dagger}\hat{\mathcal{A}}_{k}\overline{\ket{i}}\right)=O(\gamma^{w+1}) $ for $i\neq 0$.
Consequently, the bosonic AD code can approximately correct AD errors of weight up to $w$.
\hfill$\blacksquare$

\end{document}